\documentclass[twocolumn,prl,showpacs,preprintnumbers,amsmath,amssymb]{revtex4}
\usepackage{graphicx}% Include figure files
\usepackage{dcolumn}% Align table columns on decimal point
\usepackage{bm}% bold math
\usepackage{subfigure}
\begin{document}

\title{Target-skyrmions and skyrmion clusters in nanowires of chiral magnets}

\author{A. O. Leonov$^1$}
\thanks
{Corresponding author} 
\email{a.leonov@rug.nl}
\author{U. K. R\"o\ss ler$^2$}
\author{M. Mostovoy$^1$}
\affiliation{$^1$Zernike Institute for Advanced Materials, University of Groningen, Groningen, 9700AB, The Netherlands}
\affiliation{$^2$Leibniz Institute for Solid State and Materials Research, IFW, Dresden, Germany}

\begin{abstract}%
   In bulk non-centrosymmetric magnets the chiral Dzyaloshinskii-Moriya exchange stabilizes tubular skyrmions with a reversed magnetization in their centers. While the double-twist is favorable in the center of a skyrmion, it gives rise to an excess of the energy density at the outskirt. Therefore, magnetic anisotropies are required to make skyrmions more favorable than the conical spiral state in bulk materials. 
  Using Monte Carlo simulations, we show that in magnetic nanowires unusual skyrmions with a doubly twisted core and a number of concentric helicoidal undulations (target-skyrmions) are thermodynamically stable even in absence of single-ion anisotropies. Such skyrmions are free of magnetic charges and, since the angle describing the direction of magnetization at the surface depends on the radius of the nanowire and an applied magnetic field, they carry a non-integer skyrmion charge $s > 1$. This state competes with clusters of spatially separated $s=1$ skyrmions. For very small radii, the target-skyrmion transforms into a skyrmion with $s < 1$, that resembles the vortex-like state stabilized by surface-induced anisotropies.
\end{abstract}
\maketitle
\textit{1. Introduction.}
%\label{intro}
%
%1. General words about skyrmions
%
In non-centrosymmetric magnetic systems, Dzyaloshinskii-Moriya interactions (DMI) \cite{Dzyaloshinskii} stabilize two-dimensional chiral modulations (skyrmions) \cite{,Hubert,Roessler11}. 
These solitonic states may exist either as localized countable excitations or condense into multiply modulated phases - skyrmion lattices \cite{Hubert,thesis}. 
Squeezed into the spots of nanometer scale, they  have perspectives to be used in a completely new generation of spintronic and data storage devices where the skyrmionic units may be
created, manipulated, and eventually driven together to form versatile magnetic patterns \cite{Fert12}.
The chiral skyrmions are smooth, topological, and static spin textures. Therefore, they may have advantages for applications over other axisymmetric two-dimensional magnetization distributions like magnetic bubble domains \cite{domains} or vortices appearing in the magnetic nanodots due to the dipole-dipole interactions \cite{{Wachowiak02}}. 
However, because of unfavourable energetics (the double-twist is favorable only in the center of a skyrmion and gives rise to an excess of the energy density at the outskirt  \cite{Nature06}) in many bulk chiral magnets skyrmionic states usually exist as metastable states screened by lower-energy helical and conical phases. As a result the existence regions of multidimensional modulations are restricted to a close vicinity of the ordering temperatures (e.g. precursor effects in cubic helimagnets \cite{Wilhelm11}).
%
%2. Why do we need confined geometries

%.2.1. Confined geometries give rise to surface-induced interaction. In particular, in thin films, SA stabilizes skyrmions and kills conical phase

\begin{figure}
\centering
\includegraphics[width=8.8cm]{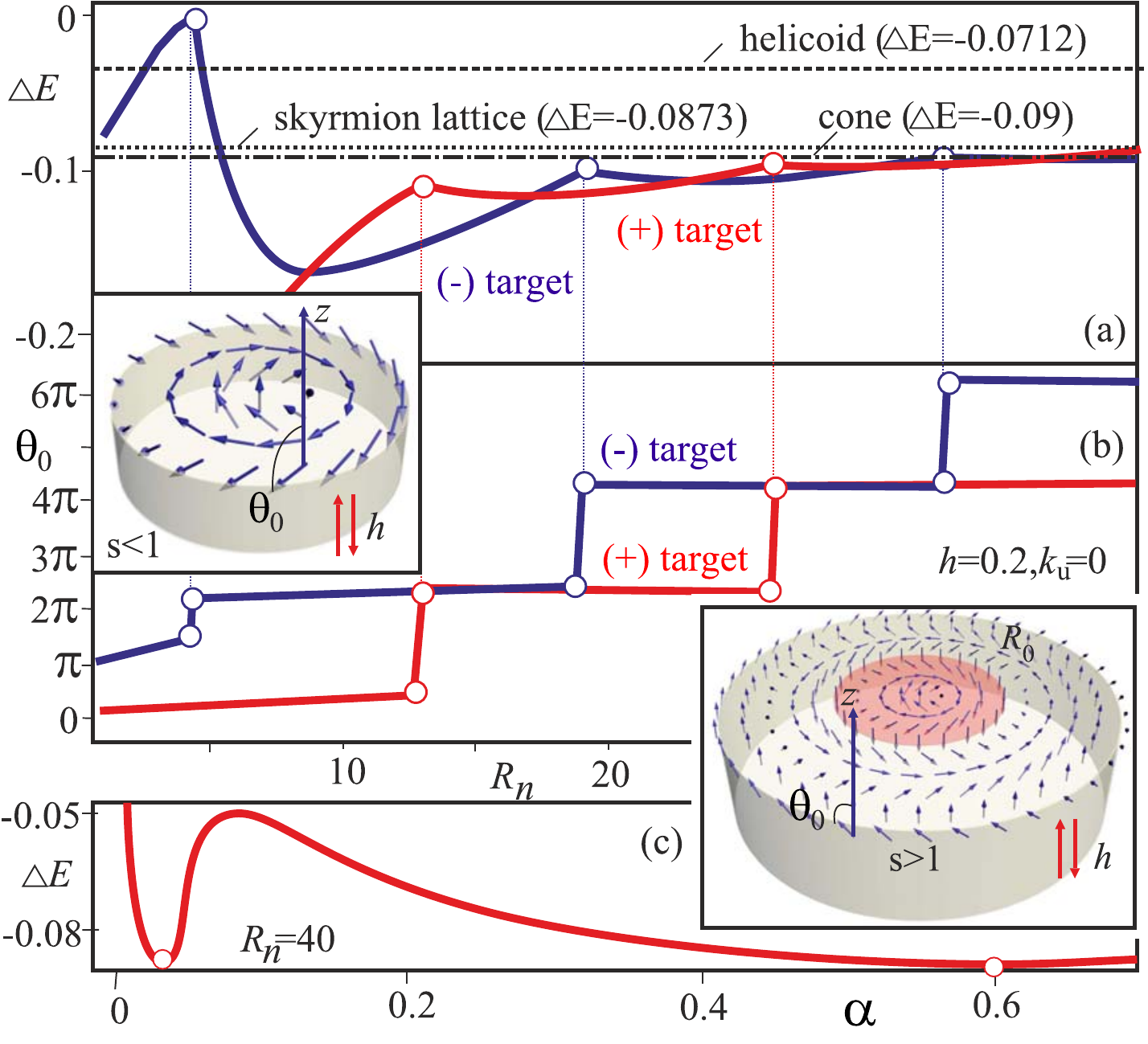}
%\vspace*{5cm}       % Give the correct figure height in cm
\caption{(Color online)  Energy densities $\Delta E$ (in dimensionless units according to Eq. (3)) counted from the energy density of the homogeneous state  (a)  and the edge angles $\theta_0$ (b) of the $(\pm)$-targets (red and blue curves, correspondingly) versus  the nanowire radius $R_n$ for $h=0.2, k_u=0$. Insets show the structure of target-skyrmions with skyrmion numbers $s<1$ and $s>1$.  Dotted, dash-and-dot, and dashed lines show the energy densities of skyrmion lattice, conical, and spiral states in a bulk helimagnet, cerrespondingly. $R_0$ is the radius of the central skyrmion. (c) Energy density of a $(+)$-target in dependence on the initial derivative $(d\theta/ d \rho)_{\rho = 0} =\alpha$.}
\label{fig1}       % Give a unique label
\end{figure}

In the \textit{confined} artificial systems on the contrary,  the existence area of skyrmionic states may be essentially extended due to surface/interface induced interactions \cite{Johnson96} producing additional stabilization effects for chiral skyrmions \cite{thesis,Butenko10}.
%
%Recently, such skyrmionic textures have been visualized by Lorentz microscopy in nanolayers of FeGe and Fe$_{0.5}$Co$_{0.5}$Si in a broad temperature range far lower the Curie point (Tc) [7]. 
%
Recently, chiral skyrmions (isolated and bound into hexagonal lattices) have been visualized by Lorentz microscopy in nanolayers of FeGe  and Fe$_{0.5}$Co$_{0.5}$Si \cite{Yu} in a broad range of magnetic fields and temperatures far below the precursor regions.
Remarkably, high-symmetry nanostructured objects (like magnetic nanowires \cite{nanowires}, nanodisks \cite{Butenko09}, or nanorings) provide not only the stabilization effect of surfaces on the skyrmion states in a nanoobject cross-section, but also introduce the problem of "skyrmion packing",  i.e.  the surface area of the cross-section defines the number of skyrmions in the cluster and their characteristic sizes.
The same effect has been recently proposed for skyrmions that lie in the plane of the MnSi-film \cite{Wilson12}.
Inspite of the larger energy due to the non-equilibrium skyrmion radii, coexistence of the skyrmion cluster with edge states in magnetic nanowires makes it energetically favourable in comparison with the  conical phase  even without additional energy contributions. 
In the present paper using Monte Carlo simulations (for details of the method see Ref. \cite{thesis}), we investigate field evolution of skyrmion clusters for different nanowire radii.

We also show that the confined cylindrical geometry opens up the perspectives to create chiral magnetic configurations that do not exist in bulk materials, namely, skyrmions with a doubly-twisted core and a number of concentric helicoidal undulations.
Such "target-skyrmions" are free of magnetic charges and, since the angle describing the direction of magnetization at the surface depends on the radius of the nanowire and an applied magnetic field, they carry a non-integer skyrmion charge $s > 1$ (see inset on the right side of Fig. \ref{fig1}). 
%
%By using the standard shooting method, 
%
We determine the equilibrium parameters of these chiral modulations as functions of applied fields, nanowire's radii, and values of surface and volume uniaxial anisotropies. 
For very small radii or for a strong applied field, the target-skyrmion transforms into a skyrmion with $s < 1$ (see inset on the left side of Fig. \ref{fig1}). %, that resembles the vortex-like state stabilized by surface-induced anisotropies \cite{Roessler02}.
We also investigate the competition of this state with clusters of spatially separated $s=1$ skyrmions (Fig. \ref{fig3}).

\textit{2. Phenomenological model.}
%\label{sec-1}
%
\begin{figure}
\centering
\includegraphics[width=8.8cm]{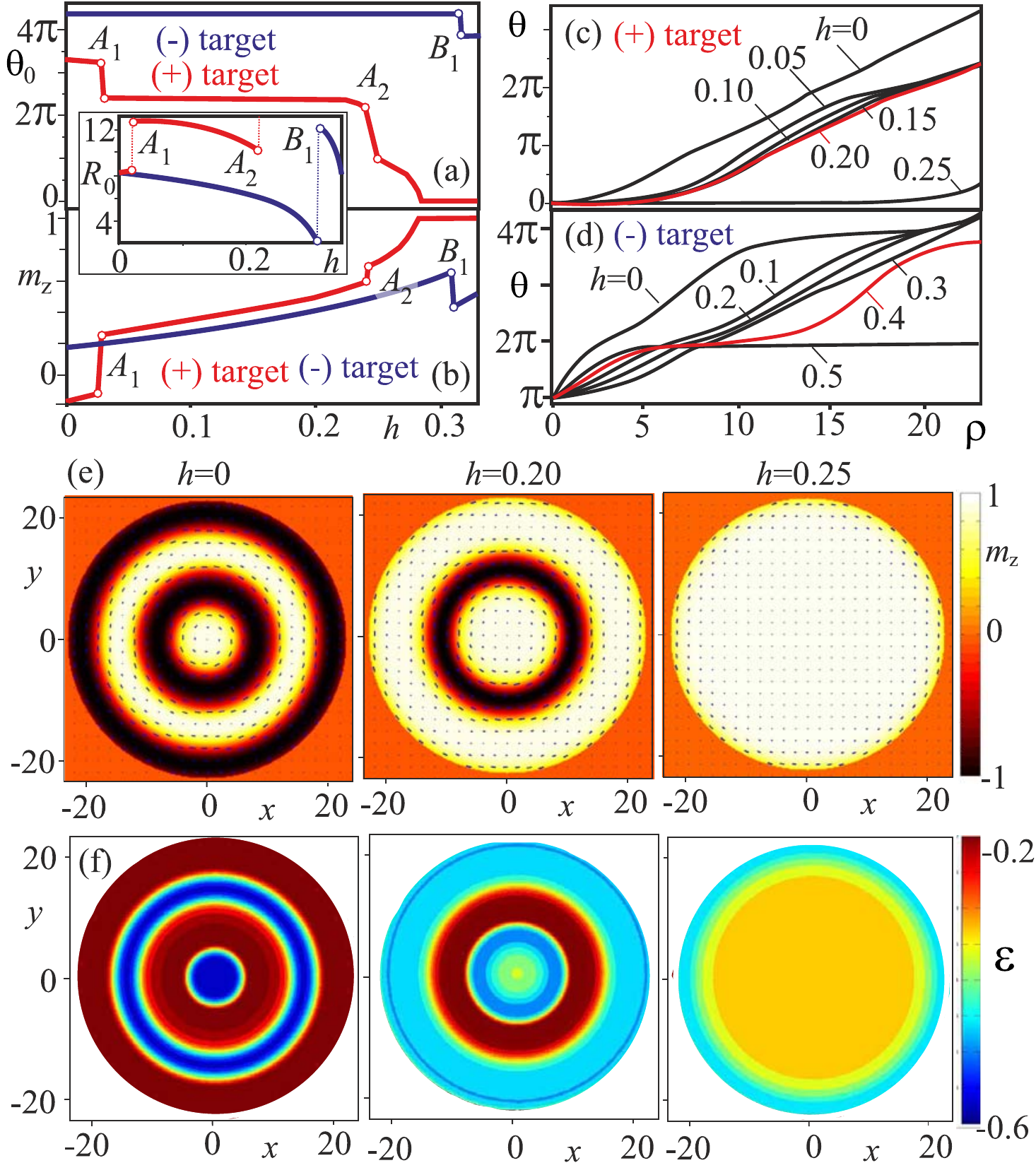}
%\vspace*{5cm}       % Give the correct figure height in cm
\caption{(Color online) Edge angles $\theta_0$ (a) and  $m_z$-components of the magnetization (b) in (+)-targets (red curves) and (-)-targets (blue lines) depending on the applied magnetic field $h$ for the fixed value of the uniaxial anisotropy, $k_{\mathrm{u}}=0.1$, and nanowire radius, $R_n=23$. Inset demonstrates the variation of the radius $R_0$ of the central skyrmion. (c) and (d) show angular profiles $\theta(\rho)$ in ($\pm$)-targets for each field interval. Contour plots in (e) show $m_z$-component of the magnetization in (+)-targets with two rings (first panel), one ring (second panel), and without rings (third panel) of the negative magnetization $m_z$, i.e. continuous unwinding of the target-skyrmion. (f) corresponding distributions of the energy density $\varepsilon$ (Eq. (\ref{density})).
\label{fig2}       % Give a unique label
}
\end{figure}
%
%In magnetic nanostructures complex physical and chemical processes on the surfaces  and interfaces essentially modify or dominate their magnetic properties compared to  corresponding bulk or {\it micro}scale objects from the same materials  \cite{Johnson96,PRL01,BogdanovJMMM02,Bode07}.
%
In many practically important cases the magnetic energy of  nanosystems can be simplified by representing it in a sum of the volume ($w_{\mathrm{v}}[\mathbf{M}(\mathbf{r})]$) and surface ($w_{\mathrm{s}}([\mathbf{M}(\mathbf{r}))$)
contributions: %$W=\int_{V}w_{\mathrm{v}}d\mathbf{x}+\int_{S}w_{\mathrm{s}}d\mathbf{x}$.
\begin{equation}
W=\int_{V}w_{\mathrm{v}}d\mathbf{r}+\int_{S}w_{\mathrm{s}}d\mathbf{r}\,.
\label{functional1}
\end{equation}
Then the equilibrium distributions of the magnetization $\mathbf{M}(\mathbf{r})$ are derived by solving the Euler equations for the volume functional $w_{\mathrm{v}}[\mathbf{M}]$ with the boundary conditions imposed by surface energy $w_{\mathrm{s}}[\mathbf{M}]$ (see e.g. \cite{domains}).
In this paper we solve such a problem for a nanowire with the circular cross-section in the $xy$-plane and infinitely extended  into $z$-direction. 
We consider a simplified model of a chiral ferromagnet  with the volume energy density
\begin{equation}
%w=\int_{V}(A\sum_{i,j}(\partial_i m_j)^2+Â \varepsilon_{ijk}m_i\delta_jm_k+w_{0})d\mathbf{r}+\int_{S}f_sd\mathbf{r}.
%w_{\mathrm{v}}=A(\mathrm{grad}\,\mathbf{m})^2+D\,\mathbf{m}\cdot \mathrm{rot}\,\mathbf{m}-\mathbf{H}\cdot\mathbf{M}-K_{\mathrm{u}}(\mathbf{m}\cdot\mathbf{a})^2,
w_{\mathrm{v}}[\mathbf{M}]=A(\nabla \mathbf{m})^2+D\,\mathbf{m}\cdot \nabla \times \mathbf{m}-\mathbf{H}\cdot\mathbf{M}
-K_{\mathrm{u}}(\mathbf{m}\cdot\mathbf{a})^2,
\label{functional2}
\end{equation}
where $A$ is a stiffness constant, $D$ is a Dzyaloshinskii constant, $\mathbf{H}$ is the applied magnetic field aligned along the nanowire axis, $\mathbf{m}$ = $\mathbf{M}/M_s$ is the unity vector along the magnetization  ($M_s = |\mathbf{M}|$),  $K_u > 0 $ is the constant of volume uniaxial anisotropy, unity vector $\mathbf{a}$ defines "easy" anisotropy direction.
%
%free-boundary conditions.
%

\textit{3. Target-skyrmions.}
To describe target-skyrmions in a nanowire of radius $R_n$, we consider axisymmetric distributions of the magnetization and express the magnetization vector  $\textbf{m}$ in terms of spherical coordinates, $\mathbf{m}=(\sin\theta\cos\psi;\sin\theta\sin\psi;\cos\theta)$, and the spatial variables in cylindrical coordinates, $\mathbf{r}=(r\cos\varphi;r\sin\varphi;z)$.
The target-skyrmions are characterized by the magnetization direction in the center, $r=0$,  along  the field ($\theta(0)=0$, (+)-targets) or opposite to the field ($\theta(0)=\pi$, (-)-targets), and by the varying angle $\theta_0$ at the surface of a nanowire (Fig. \ref{fig1}).   
%
%The  variational problem for functional  (\ref{density}) has rotationally symmetric solutions $\psi=\varphi\pm\pi/2$, $\theta = \theta(\rho)$.
%
By substituting the solution for $\psi=\varphi+\pi/2$ into Eq.~(\ref{functional2}) and integrating with respect to $\varphi$ the  energy density of target-skyrmions 
%with respect to the homogeneous state 
can be reduced to the following form $E=(2/R_n^2)\int^{R_n}_{0} \varepsilon(\rho)\rho d \rho$, where
%
 %\begin{eqnarray}
%\varepsilon(\rho)&=& A \left[\left(\frac{d \theta}{d \rho}\right)^{2}+\frac{1}{\rho ^2}\sin^2\theta \right]
%+K \cos^2\theta \label{density}
%\\ \nonumber
%&-&HM_s\cos \theta-D \left(\frac{d \theta}{d \rho}+\frac{1}{\rho}\cos\theta\sin\theta\right),
%\end{eqnarray}
%
% \begin{eqnarray}
%\varepsilon(\rho)&=& \left[\left(\frac{d \theta}{d \rho}\right)^{2}+\frac{1}{\rho ^2}\sin^2\theta \right]
%+k_u \cos^2\theta \label{density}
%\\ \nonumber
%&-&h\cos \theta-D \left(\frac{d \theta}{d \rho}+\frac{1}{\rho}\cos\theta\sin\theta\right),
%\end{eqnarray}
%
\begin{equation}
%\varepsilon(\rho)=&\left[\left(\frac{d \theta}{d \rho}\right)^{2}+\frac{1}{\rho ^2}\sin^2\theta \right]+k_u \sin^2\theta+h(1-\cos \theta)-\nonumber\\
%&+\left(\frac{d \theta}{d \rho}+\frac{1}{\rho}\cos\theta\sin\theta\right), 
%\varepsilon(\rho)=&\left[\left(\frac{d \theta}{d \rho}\right)^{2}+\frac{1}{\rho ^2}\sin^2\theta \right]-k_u \cos^2\theta-h\cos \theta+\nonumber\\
%&+\left(\frac{d \theta}{d \rho}+\frac{1}{\rho}\cos\theta\sin\theta\right), 
\varepsilon(\rho)=\theta_{\rho\rho}^{2}+\frac{\sin^2\theta}{\rho ^2}-k_u \cos^2\theta-h\cos \theta+\theta_{\rho}+\frac{\sin(2\theta)}{2\rho}, 
\label{density}
\end{equation}
and non-dimensional field $h=HM_sA/D^2$, anisotropy $k_{\mathrm{u}}=K_uA/D^2$, and coordinate $\rho=rD/A$ have been introduced, $\theta_{\rho\rho}=d^2\theta/d\rho^2, \theta_{\rho}=d\theta/d\rho$. %Magnetic field $h$ is assumed to be aligned along the nanowire axis.

First, we consider magnetic nanowires without any constrained conditions at the edge.  
By solving an  \textit{auxiliary} Cauchy problem  with initial conditions $\theta(0)=0,\,\pi, \quad  (d\theta/ d \rho)_{\rho = 0} =\alpha$ \cite{Butenko09}, one finds that in the applied magnetic field the energy density has several minima  for particular values of $\alpha$ satisfying the natural boundary conditions, $(d\theta/ d \rho)_{\rho = R_n} =0.5$ \cite{Du13}.
Fig. \ref{fig1} (c) has been plotted as an example for $\theta(0)=0$.
Solutions for target-skyrmions in the minima differ by $2\pi$ of the edge angle $\theta_0$ and are energetically degenerate for particular values of the nanowire radii (i.e. at cusp points of energy densities of Fig. \ref{fig1} (a) corresponding to jumps of $\theta_0$ in Fig. \ref{fig1} (b)).
In general, the nanowire radius $R_n$ allows the number of rings in the target (i.e. skyrmion number $s$) and the radius of the central skyrmion $R_{\mathrm{0}}$ to vary and makes energetically favourable either (+)- or (-)-targets (blue and red curves in Fig. \ref{fig1} (a)) for a fixed values of the field and anisotropy constant.
In Fig. \ref{fig1} (a) plotted for $h=0.2, k_{\mathrm{u}}=0$ the energy of $(\pm)$-targets is the lowest for a wide range of radii as compared with the conical phase ($\Delta E=(-1+2h+4k_{\mathrm{u}})/(-4+16k_{\mathrm{u}})$), helicoid, and skyrmion lattice in the bulk materials.

%For two-column wide figures use syntax of figure~\ref{fig-2}

In an applied magnetic field any (+)-target may be unwound into homogeneously magnetized state (red lines in Fig. \ref{fig2} (a), (b)).
Such a process is accompanied by the jumps of the edge angle $\theta_0$ (Fig. \ref{fig2} (a)) and  the magnetization $m_z$ in the magnetization curves (Fig. \ref{fig2} (b)),  which are the result of the successive disappearance of the spiral rings in the (+)-targets. 
Note, that in the field $h(A_1)$ the jump occurs from the negative value of the magnetization $m_z$ (target-skyrmion has two rings with the negative $m_z$-component, Fig. \ref{fig2} (e), $h=0$) to the positive one (one ring, Fig. \ref{fig2} (e), $h=0.20$).
For $h>h(A_2)$ only target-skyrmion with $s<1$ can exist (Fig. \ref{fig2} (e), $h=0.25$), which transform into the fully magnetized homogeneous state in very large magnetic fields.
The reason for this is the ring of the negative energy density due to the deflection of the magnetization (Fig. \ref{fig2} (f), $h=0.25$), which surrounds the disc with the energy density as in a homogeneous state. % (Fig. \ref{fig2} (f), $h=0.25$). 
$(-)$-target in the applied magnetic field (blue lines in Fig. \ref{fig2} (a), (b)) may be unwound up to the single skyrmion with $s=1$ in the center of a nanowire. 
For both target-skyrmions the radius $R_0$ of the central skyrmion (see inset of Fig. \ref{fig2} (a),(b) also changes discontinuously as contrasted to the gradual increase of the skyrmion radius in the hexagonal lattice of bulk materials \cite{Hubert,Butenko10,thesis}.
Angular profiles $\theta(\rho)$  for ($\pm$)-targets (Fig. \ref{fig2} (c), (d)) with broadened parts along the applied magnetic field show the solutions in each field interval. %broadening parts along the field and transition 

%Energy density $\varepsilon$ in such a state has the same value as for homogeneous state in the central disk and is surrounded by the ring with negative 

%For a fixed value of the field $h=0.2$ and anisotropy constant $k_{\mathrm{u}}=0$, either (+)- or (-)-targets may be favourable (blue and red curves in Fig. \ref{fig1} (a)) for variable nanowire radii.
% 
%In the applied magnetic field there are two enrgy minima with the difference $2\pi$ between the edge angles $\theta_0$.

%In minimizing Eq. (\ref{density}) we consider an  \textit{auxiliary} Cauchy problem  with initial conditions $\theta(0)=0, \quad  (d\theta/ d \rho)_{\rho = 0} =\alpha$ for different values of $\alpha>0$.
%
%In the applied magnetic field 

%The Euler equation for $\theta(\rho)$ obtained by minimizing (\ref{density}) with the natural boundary conditions , $\theta(0)=0,\, \quad \left(d \theta / d \rho \right)_{\rho =R_d}= 0.5$ yield the equilibrium solutions for target-skyrmions.

\begin{figure}
\centering
\includegraphics[width=8.8cm]{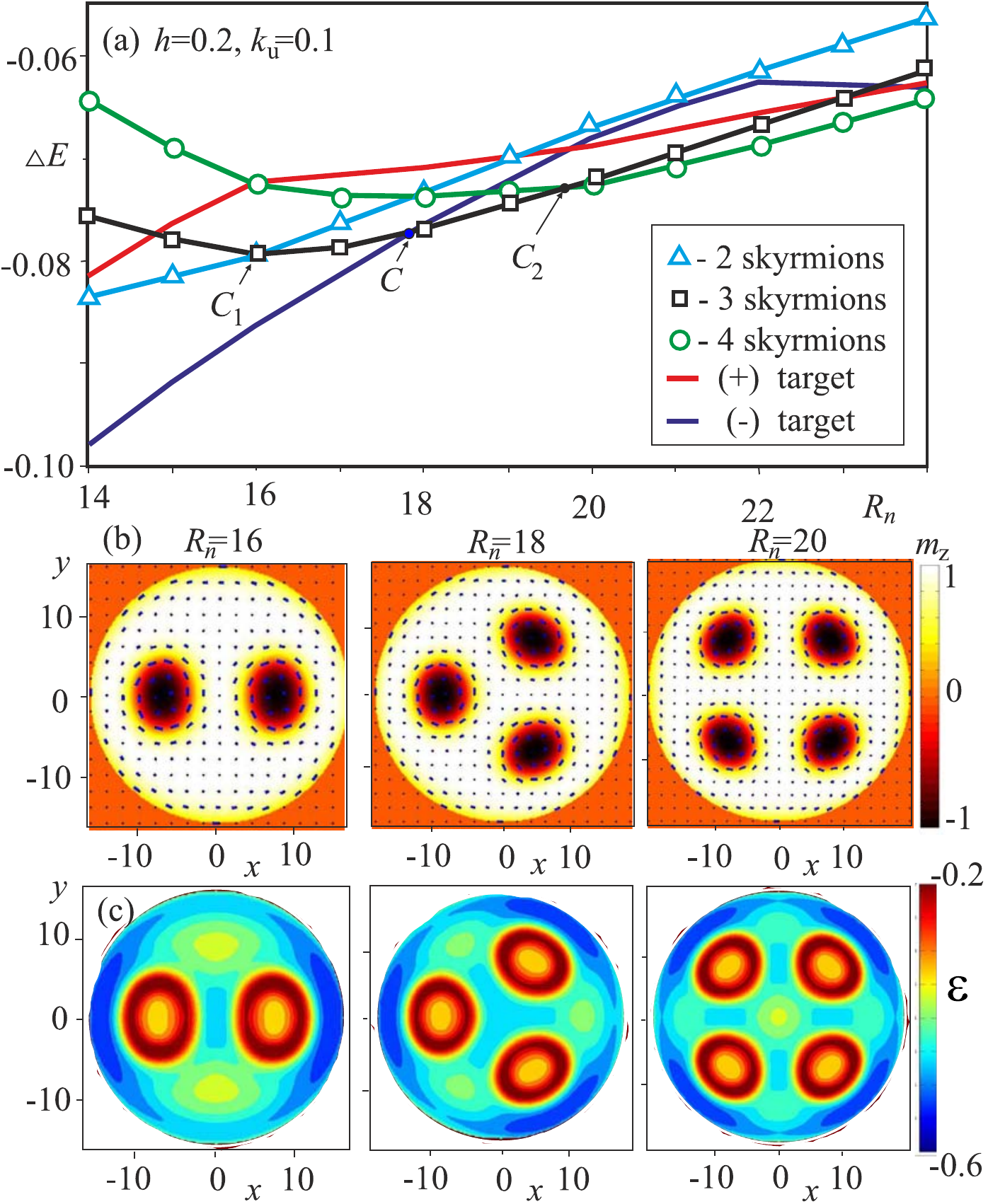}
%\vspace*{5cm}       % Give the correct figure height in cm
\caption{(Color online) (a) Energy densities $\Delta E$ with respect to the homogeneous state of ($\pm$)-targets (red and blue curves) and skyrmion clusters composed from two (triangles), three (squares), and four (circles) skyrmions for $h=0.2, k_{\mathrm{u}}=0.1$.  Contour plots in (b) and (c) show $m_z$-component of the magnetization in the skyrmion clusters and corresponding distributions of the energy density $\varepsilon$ (Eq. (\ref{density})).}
\label{fig3}       % Give a unique label
\end{figure}

%\textit{Unwinding of the target in an applied magnetic field.}

\textit{4. Clusters of skyrmions in a nanowire cross-section.}
Clusters of spatially separated ($s=1$)- skyrmions may become thermodynamically stable for some $R_{\mathrm{n}}$ and $k_{\mathrm{u}}$. % in a certain interval of nanowire radii and constants of uniaxial anistropy.
As an example in Fig. \ref{fig3} (a), the energy densities of ($\pm$)-targets (blue and red lines) for fixed values of the field $h=0.2$ and uniaxial anisotropy $k_{\mathrm{u}}=0.1$ have been plotted together with the energy densities of skyrmion clusters composed from two, three, and four skyrmions.
%
%These data not only show the energy advantage of clusters for larger nanowire radii (for $h=0, k_{\mathrm{u}}=0$ the critical radius is $R_n=61.8$), but also indicate that the system may be degenerate with different skyrmion numbers at certain points $C_1$ and $C_2$. 
%
These data show the energy advantage of skyrmion clusters for nanowire radii larger than some critical value: for $h=0.2, k_{\mathrm{u}}=0.1$ the critical radius is $R_n=17.9$ (point $C$ in Fig. \ref{fig3} (a)).
The data also indicate that the system may be degenerate with different skyrmion numbers at some certain points $C_1$ and $C_2$. 
For $R_{\mathrm{n}}<R_{\mathrm{n}}(C_1)$ a two-skyrmion cluster has the lowest energy and is replaced by the three-skyrmion cluster in the interval $R_{\mathrm{n}}(C_1)<R_{\mathrm{n}}<R_{\mathrm{n}}(C_2)$ and by the four-skyrmion cluster for $R_{\mathrm{n}}>R_{\mathrm{n}}(C_2)$.
Radii of single skyrmions in the clusters do not coincide with those in the skyrmion lattice of bulk materials which is known to lead to an increase of their energy density \cite{Hubert,Butenko10,thesis}. 
However, coexistence of skyrmion clusters with the edge states, like those with the deflected magnetization shown in Fig. \ref{fig2} (e) for $h=0.25$, lowers the cluster energy (the energy density of the skyrmion lattice in the bulk materials for the same values of $h$ and $k_{\mathrm{u}}$ is $\Delta E=-0.045$). 
Fig. \ref{fig3} (c) shows distributions of energy density in all clusters  with negative energy contributions from the nanowire edge. 
Fig. \ref{fig3} (b) shows contour plots of $m_z$-components of the magnetization in the clusters. 
Edge states still keep the clusters free of magnetic charges, since the magnetization is parallel to the nanowire surface.

\begin{figure}
\centering
\includegraphics[width=8.8cm]{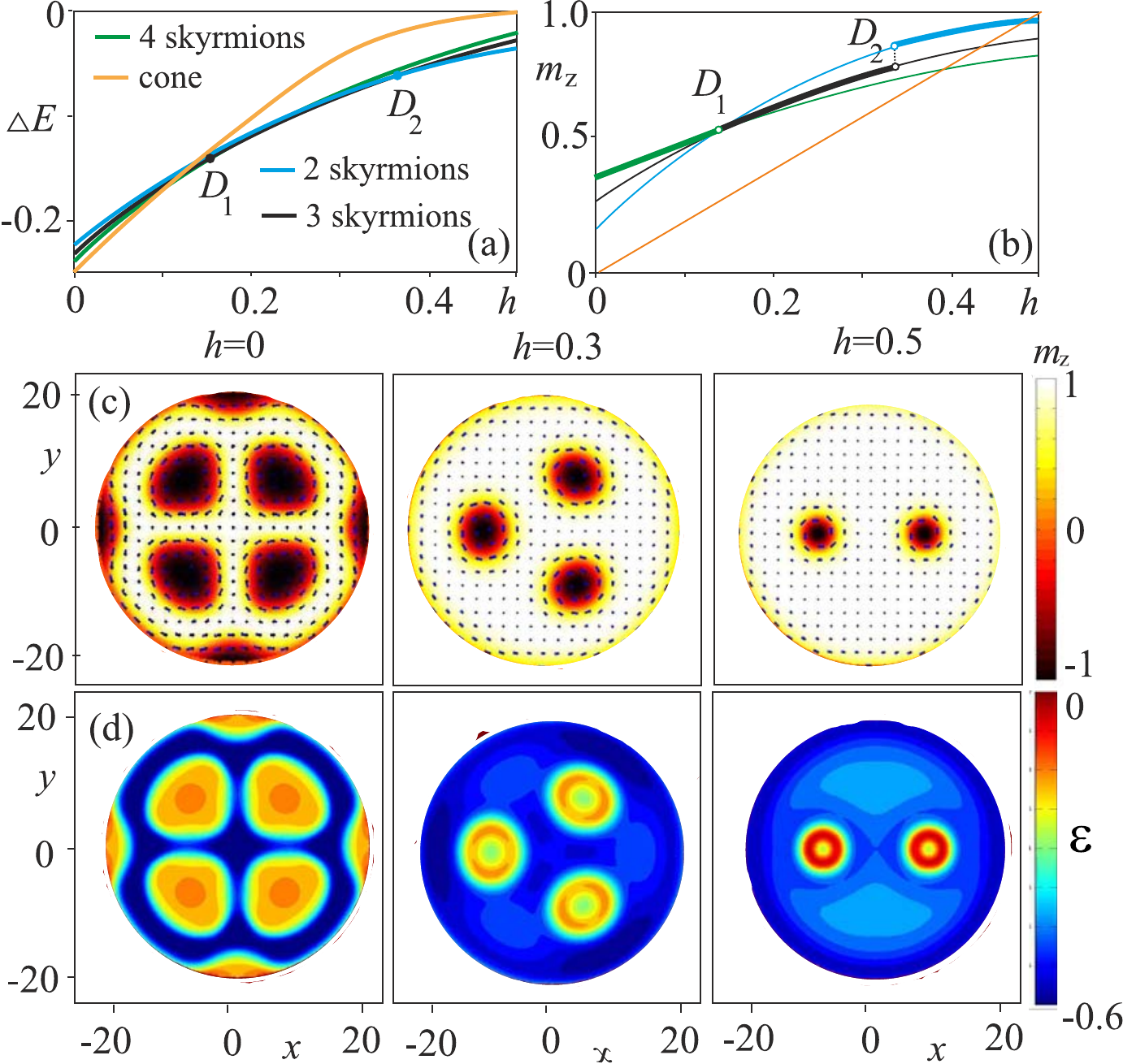}
%\vspace*{5cm}       % Give the correct figure height in cm
\caption{(Color online)  Energy densities $\Delta E$ (a) with respect to the homogeneous state and normalized magnetization $m_z$ (b) of skyrmion clusters and conical phase (orange line) in dependence on the applied magnetic field $h$ for the fixed nanowire radius $R_n=20$ and $k_{\mathrm{u}}=0$. (c) and (d) show contour plots of $m_z$-component and energy density $\varepsilon$ (Eq. (\ref{density}) in the skyrmion clusters for different values of the field.}
\label{fig4}       % Give a unique label 
\end{figure}

The system may also become degenerate in magnetic fields. 
Fig. \ref{fig4} (a) shows energy densities of skyrmion clusters for $R_n=20,\, k_{\mathrm{u}}=0$ with the four-skyrmion cluster having the minimal energy for $h<h(D_1)$, three-skyrmion cluster - for $h(D_1)<h<h(D_2)$, and two-skyrmion cluster - for $h(D_2)<h$. 
Such a discrete change of the skyrmion number is related to the repulsive interaction between skyrmions.
Having no possibility to change the lattice period continuously as in the bulk materials \cite{Hubert,thesis,Butenko10}, skyrmions in clusters are "squeezed" by their interactions.
The edge state with the negative energy density, however, may prevent a continuous deformation of the skyrmions by keeping the skyrmion number constant.
In a strong magnetic field skyrmions become very localized (Fig. \ref{fig4}). 
Their coexistence with the edge state makes the critical field, where skyrmions disappear, much larger than the corresponding saturation field for the bulk.
In low fields the skyrmions undergo elliptical transformation and may elongate into spiral states. 
Note that the conical phase (the orange line in Fig. \ref{fig4} (a)) within the isotropic model has the lowest energy only for low magnetic fields, that is the effect of "skyrmion packing" and coexistence with edge states make skyrmion clusters thermodynamically stable.

\textit{5. Vortices stabilized by the surface-induced uniaxial anisotropy.}
Target-skyrmions with a skyrmion number $s<1$ may be also stabilized by the induced magnetic anisotropy from lateral surfaces \cite{Roessler02} even without DMI. 
Reduced coordination number at the surfaces causes strong magnetic surface anisotropy as known for planar surfaces and ultrathin films \cite{Neel}. 
For nanowires, these effects may become very prominent and can be amplified by the curved surface structure.
The surface contribution in Eq. (\ref{functional1}) may be written as $w_{\mathrm{s}}[\mathbf{M}]=-K_s\cos^2\theta$, where $K_s$ is the constant of surface-induced uniaxial anisotropy. 
According to depth-resolving experimental techniques the surface-induced interactions may spread into the depth of the magnetic nanosystems \cite{nanowires2} (i.e. giving the contribution to $w_{\mathrm{v}}[\mathbf{M}]$ in Eq. (\ref{functional1})) and favours magnetization orientations either perpendicular or parallel to the surface.
Therefore in magnetic nanowires, the magnetization vector may rotate either along the radial direction (N\'eel vortices) or perpendicular to it (Bloch vortices) and may have both chiralities.
Phenomenological theory of these vortices in magnetic nanowires and nanotubes have been developed in Refs. \cite{Roessler02,thesis}.
%
%Noncollinear remanent states have been reported in Fe nanowires, which lead to high saturation fields in fields applied perpendicularly to the nanowire \cite{Zhang03}. Anisotropic magnetoresistance measurements performed on nanowires embedded in the polymer membrane have been used as a probe of the magnetization orientation with respect to the current \cite{Rheem07}. In Ni nanowires vortex-like non-collinear states are probably the reason of slight deviations of magnetoresistance from constant value in magnetic field along the nanowire axis \cite{Rheem07}. The analysis of substructures in the ferromagnetic-resonance (FMR) lines of Ni nanowire indicate the presence of rather strong surface-anisotropies \cite{Ebels01}.

\textit{6. Conclusions.}
Using micromagnetic approach and Monte-Carlo simulations for a basic micromagnetic Dzyaloshinskii model of cylindrical nanowires, we analyze the internal structure and competition of target-skyrmions and skyrmion clusters.
Clusters of few $(s=1)$- skyrmions with two degenerate characteristic sizes form a global energy minimum only for a nanowire radius exceeding some critical value. 
In wires with smaller radii the ground-states are target-skyrmions with $s<1$, which resemble vortices stabilized by the surface-induced uniaxial anisotropy. 
In the applied magnetic field target-skyrmions may be unwound into homogeneous state, skyrmion clusters continuously decrease their skyrmion number.

The authors are thankful to S. Bl\"ugel, A. N. Bogdanov, T. Monchesky, R. Stamps for useful discussions. This study was supported by the Stichting voor Fundamenteel Onderzoek der Materie (FOM).

%
% BibTeX or Biber users please use (the style is already called in the class, ensure that the "woc.bst" style is in your local directory)
% \bibliography{name or your bibliography database}

\end{document}